\documentclass[reprint,superscriptaddress,floatfix,amsmath,amssymb,pra,aps,showpacs,twocolumn]{revtex4-2}
\usepackage{amsmath,amsfonts,amssymb,amsthm,graphics,graphicx,epsfig,bbm}
\usepackage[colorlinks=true,citecolor=blue,linkcolor=blue,urlcolor=blue]{hyperref}
\usepackage[usenames]{color}
\usepackage{graphicx}
\usepackage[export]{adjustbox}
\usepackage{subfigure}
\usepackage{amsmath}
\usepackage{epsfig}
\usepackage{dcolumn}
\usepackage{bm}
\usepackage{color}
\usepackage{times}
\usepackage{epstopdf}
\usepackage{amssymb}
\usepackage{amstext}
\usepackage{latexsym}
\usepackage{amsfonts}
\usepackage{psfrag}
\usepackage[utf8]{inputenc}
\usepackage{graphicx}
\usepackage{amsmath}
\usepackage{amssymb}
\usepackage{dsfont}
\usepackage{tensor}
\usepackage{color}

\usepackage{upgreek}
\usepackage{braket}
\usepackage{algpseudocode}
\usepackage{algorithm}
\usepackage{orcidlink} 
\usepackage[normalem]{ulem}

\UseRawInputEncoding

\renewcommand{\Re}{\mathrm{Re}}
\renewcommand{\Im}{\mathrm{Im}}

\newcommand{\abs}[1]{| #1 |}

\begin{document}

\title{Learning spectral density functions in open quantum systems}

\author{Felipe Peleteiro\orcidlink{0009-0007-3711-2304}}
\affiliation{S\~{a}o Paulo State University (UNESP), School of Sciences, 17033-360, Bauru, SP, Brazil}

\author{Jo\~{a}o Victor Shiguetsugo Kawanami Lima\orcidlink{0009-0000-0633-5213}}
\affiliation{S\~{a}o Paulo State University (UNESP), School of Sciences, 17033-360, Bauru, SP, Brazil}

\author{Pedro Marcelo Prado\orcidlink{0000-0002-8165-3817}}
\affiliation{S\~{a}o Paulo State University (UNESP), School of Sciences, 17033-360, Bauru, SP, Brazil}

\author{Felipe Fernandes Fanchini\orcidlink{0000-0003-3297-905X}}
\affiliation{S\~{a}o Paulo State University (UNESP), School of Sciences, 17033-360, Bauru, SP, Brazil}

\author{Ariel Norambuena\orcidlink{0000-0001-9496-8765}}
\email{ariel.norambuena@usm.cl}
\affiliation{Departamento de F\'isica, Universidad T\'ecnica Federico Santa Mar\'ia, Casilla 110 V, Valpara\'iso, Chile}

\date{\today}

\begin{abstract}
Spectral density functions quantify how environmental modes couple to quantum systems and govern their open dynamics. Inferring such frequency-dependent functions from time-domain measurements is an ill-conditioned inverse problem. Here, we use exactly solvable spin-boson models with pure-dephasing and amplitude-damping channels to reconstruct spectral density functions from noisy simulated data. First, we introduce a parameter estimation approach based on machine learning regressors to infer Lorentzian and Ohmic-like spectral density parameters, quantifying robustness to noise. Second, we show that a cosine transform inversion yields a physics-consistent spectral prior estimation, which is refined by a constrained neural network enforcing positivity and correct asymptotic behaviour. Our neural network framework robustly reconstructs structured spectral densities by filtering simulated noisy signals and learning general functional dependencies. 
\end{abstract}

\maketitle

\section{Introduction}

In the theory of open quantum systems, the microscopic description of the environment introduces bath frequencies $\omega_k$ and the corresponding system-bath coupling constants $g_k \in \mathbb{C}$, where $k$ may denote a discrete or continuous index. The spectral density function (SDF) is defined as $J(\omega) = \sum_k |g_k|^2\,\delta(\omega-\omega_k)$~\cite{Vega2017, Breuerbook}, which encodes how environmental modes couple to the system under interest. The Dirac delta function \(\delta(\omega - \omega_k)\) effectively models the environment as a continuous spectral distribution, facilitating analytical treatments for damping rates and time evolution. By construction, \(J(\omega) \geq 0\), and the normalization condition \(\int_{0}^{\infty} J(\omega)\,d\omega = \sum_k |g_k|^2\) must hold.

At low frequencies ($\omega\!\to\!0$), density of states arguments or noise-correlation arguments typically lead to a power-law scaling $J(\omega) \propto \omega^s$ with $s \in \mathbb{R}^{+}$. Conversely, at high frequencies ($\omega\!\to\!\infty$), the SDF must decay, reflecting the suppression of high-energy bath excitations. The most widely used phenomenological model is the Ohmic-like SDF $J(\omega)= \alpha \,\omega_c^{1-s}\,\omega^s\, e^{-\omega/\omega_c}$, 
which captures the main physical features of a broad class of environments~\cite{Caldeira1983, Leggett1985, Weiss2008}. Here, $\alpha>0$ sets the coupling strength, $\omega_c$ is the cutoff frequency, and $s$ distinguishes sub-Ohmic ($s<1$), Ohmic ($s=1$), and super-Ohmic ($s>1$) cases. While the Ohmic class enables analytical progress and provides valuable intuition, it fails to capture the complexity of structured environments, such as quantum dots~\cite{Wilson-Rae2002} or solid-state spin defects in diamond~\cite{Alkauskas2014,Norambuena2016,Cambria2023}.

Moving beyond phenomenological models, one may use \textit{ab initio} calculations~\cite{Alkauskas2014,Cambria2023} or molecular dynamics simulations~\cite{Norambuena2016,Norambuena2023}. However, these approaches are computationally demanding, susceptible to discretization artifacts, and sensitive to many-body correlations. Alternatively, microscopic arguments within exactly solvable models can be invoked, although such methods remain restricted to specific scenarios, for instance, light-matter interactions in the electric-dipole approximation~\cite{Vega2017}. 

A promising direction is to estimate \(J(\omega)\) directly from experimental data by combining artificial intelligence (AI) with physically consistent quantum models. Given a set of measurements, one may infer the SDF that best explains the observed dynamics under suitable constraints. Current AI-assisted strategies primarily employ neural networks to classify~\cite{Barr2024,Barr2025, Barr2025_arxiv} or fit parameters~\cite{Barr2025} within the Ohmic class for the pure-dephasing spin-boson model. However, a systematic AI framework capable of reconstructing structured environments beyond Ohmic-like parametrizations from noisy data remains lacking. Moreover, robustness analyses indicate that small deviations in the SDF can lead to exponentially amplified errors in expectation values, highlighting the intrinsic sensitivity of this inverse quantum problem~\cite{Plenio2017}.

In this work, we present two complementary approaches to learn SDFs from time-domain data in exactly solvable spin-boson models, which we use to benchmark our results. The first performs AI-based parameter estimation within phenomenological spectral-density families using machine-learning regressors. The second exploits a cosine transform structure to construct a nonparametric spectral prior, which is subsequently stabilized by a physics-constrained neural network. We benchmark accuracy and robustness across noise regimes and discuss how these tools extend to broader classes of open quantum systems with known models.

\begin{figure*}[ht!]
\centering
\includegraphics[width=0.9\linewidth]{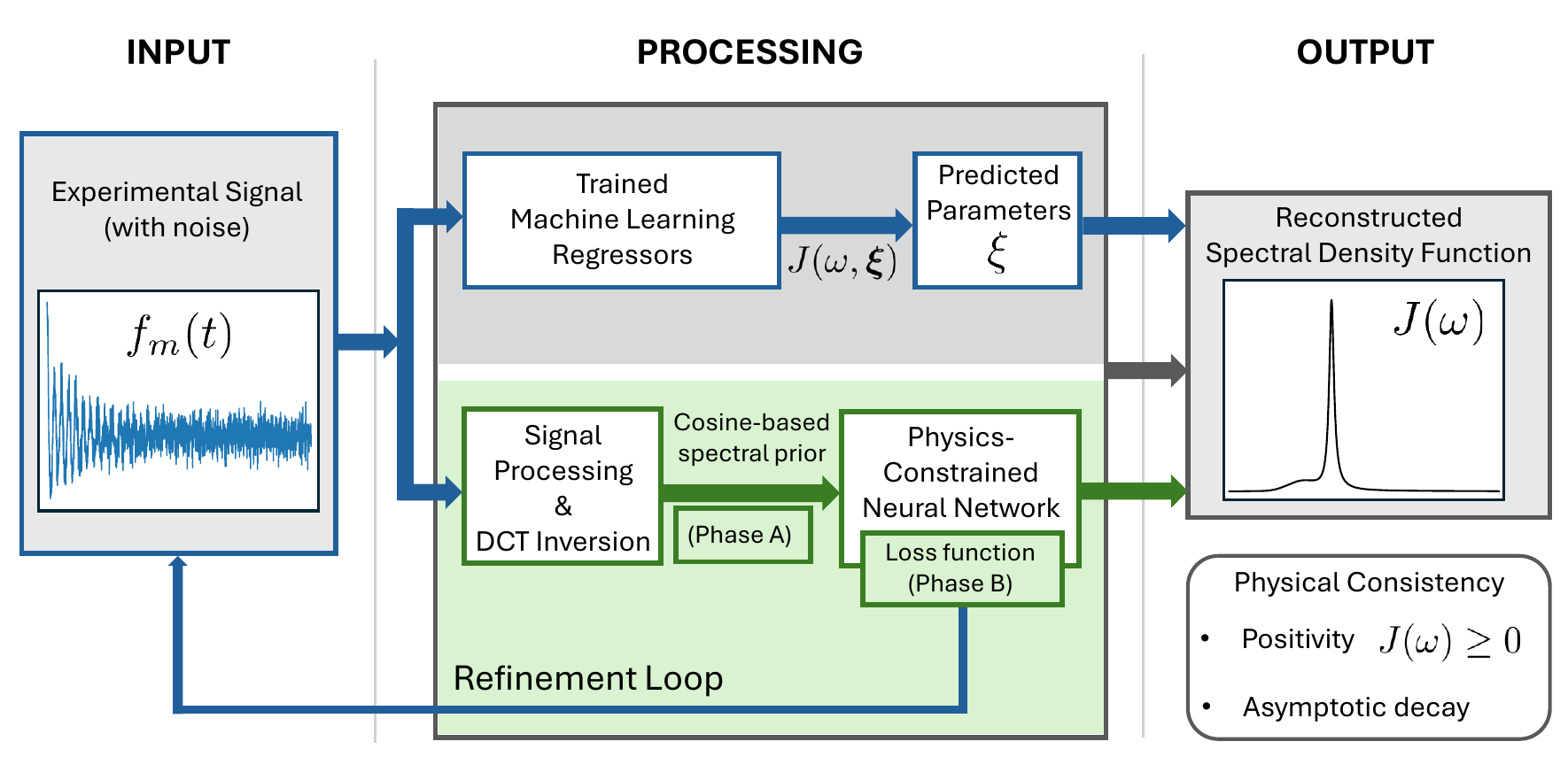}
\caption{Conceptual overview of the spectral density reconstruction problem. A quantum system coupled to a structured environment is characterized by a spectral density function $J(\omega)$, which determines measurable dynamical signals $f_m(t)$ through a known map $J(\omega) \mapsto f_m(t)$. The inverse problem, inferring $J(\omega)$ from noisy time-domain data, is ill-conditioned under finite time sampling and measurement noise. We address this challenge via two complementary AI-based strategies: (i) a parametric regression scheme for estimating SDF parameters, and (ii) a nonparametric approach combining discrete cosine transform (DCT) inversion with a constrained neural network refinement to obtain physically consistent spectral reconstructions.}
\label{Scheme}
\end{figure*}

\section{Models}
First, we consider the pure dephasing (PD) spin-boson model $\hat{H}_{\rm PD} = \hat{H}_s + \hat{H}_b +\hat{V}_{\rm PD}$, where $\hat{H}_s = (\omega_{0}/2)\hat{\sigma}_z \otimes \mathds{1}_b$ describes a two-level system with Bohr frequency $\omega_0$ and $\hat{H}_{b} = \mathds{1}_s \otimes \sum_k \omega_k \hat{a}_k^{\dagger}\hat{a}_k$ is the bosonic bath. Here, $\hat{\sigma}_z =|1\rangle \langle 1|-|0\rangle\langle 0|$ is the Pauli matrix in the two-level-system basis $\{|0\rangle,|1\rangle\}$, and $\hat{a}_k$ ($\hat{a}_k^{\dagger}$) is the annihilation (creation) boson operator. The interaction Hamiltonian for the PD channel is given by ($\hbar = 1$)

\begin{eqnarray}
    \hat{V}_{\rm PD} = \hat{\sigma}_z \otimes \sum_k\!\left(g_k \hat{a}_k + g_k^{\ast}\hat{a}_k^{\dagger}\right),
\end{eqnarray}
 where $g_k$ are the coupling constants and the SDF is defined as 
 \begin{equation} 
\label{SDF} 
J(\omega) = \sum_k |g_k|^2\,\delta(\omega-\omega_k). 
\end{equation}
The PD model captures the orbital-phonon dynamics in color centers~\cite{Ariel2020} and time-invariant quantum discord~\cite{Maniscalco2013}, both of which describe Markovian and non-Markovian regimes. For a thermal bath state $\hat{\sigma}_b = e^{-\beta \hat{H}_b}/\mathrm{Tr}[e^{-\beta \hat{H}_b}]$ with $\beta = (k_B T)^{-1}$ the inverse temperature, the exact time-local dynamics is $\dot{\hat{\rho}}_{\rm PD} = (\gamma_{\rm PD}(t)/2)\!\left[\hat{\sigma}_z \hat{\rho}_{\rm PD} \hat{\sigma}_z - \hat{\rho}_{\rm PD}\right]$~\cite{Maniscalco2013,Luczka1990}. Consistently with the convention below, the dephasing rate is
\begin{equation}
\gamma_{\rm PD}(t)=\dot G(t)
=\int_{0}^{\infty}\frac{J(\omega)}{\omega}
\coth\!\left(\frac{\beta\omega}{2}\right)\sin(\omega t)\,d\omega .
\end{equation}
Here, $n(\omega)=[e^{\beta \hbar \omega}-1]^{-1}$ is the mean boson number and $\coth(\beta\omega/2)= 2n(\omega)+1$ is the total occupation factor for absorption and emission processes. Since populations remain constant in the PD model, the quantum coherence $C(t)=\sum_{i\neq j}\!\abs{\rho_{ij}(t)}$~\cite{Baumgratz2014} is a natural observable. Experimentally, it can be reconstructed as

\begin{eqnarray}
    f_{\rm PD}(t) &=& \sqrt{\langle \hat{\sigma}_x(t)\rangle^2+\langle \hat{\sigma}_y(t)\rangle^2} = 2|\rho_{01}(0)|e^{-G(t)}, \label{PD_signal}\\
G(t) &=& \int_{0}^{\infty}\frac{J(\omega)}{\omega^2}\,\mathrm{coth}\!\left(\frac{\beta \omega}{2}\right)\!\big[1-\cos(\omega t)\big]\,d\omega,
\end{eqnarray} 
where $f_{\rm PD}(t) = C(t)$ is the quantum coherence signal for the PD model, with initial value $C(0) = 2|\rho_{01}(0)|$. Second, we consider the amplitude-damping (AD) spin-boson model, whose Hamiltonian is $\hat{H}_{\rm AD} = \hat{H}_s + \hat{H}_{b}+ \hat{V}_{\rm AD}$, in which the system and bath Hamiltonians are the same as the PD model. In the rotating-wave approximation, the system-environment interaction Hamiltonian for the AD channel is given by ($\hbar = 1$)
\begin{equation}
     \hat{V}_{\rm AD} = \sum_k\!\left(g_k \hat{\sigma}_+ \otimes \hat{a}_k + g_k^{\ast}\hat{\sigma}_- \otimes \hat{a}_k^{\dagger}\right),
\end{equation}
where $\hat{\sigma}_+=(\hat{\sigma}_x+i\hat{\sigma}_y)/2$ and $\hat{\sigma}_- = (\hat{\sigma}_+)^{\dagger}$ describe excitation and relaxation of the two-level system, respectively.

\begin{figure*}[ht!]
\centering
\includegraphics[width=1\linewidth]{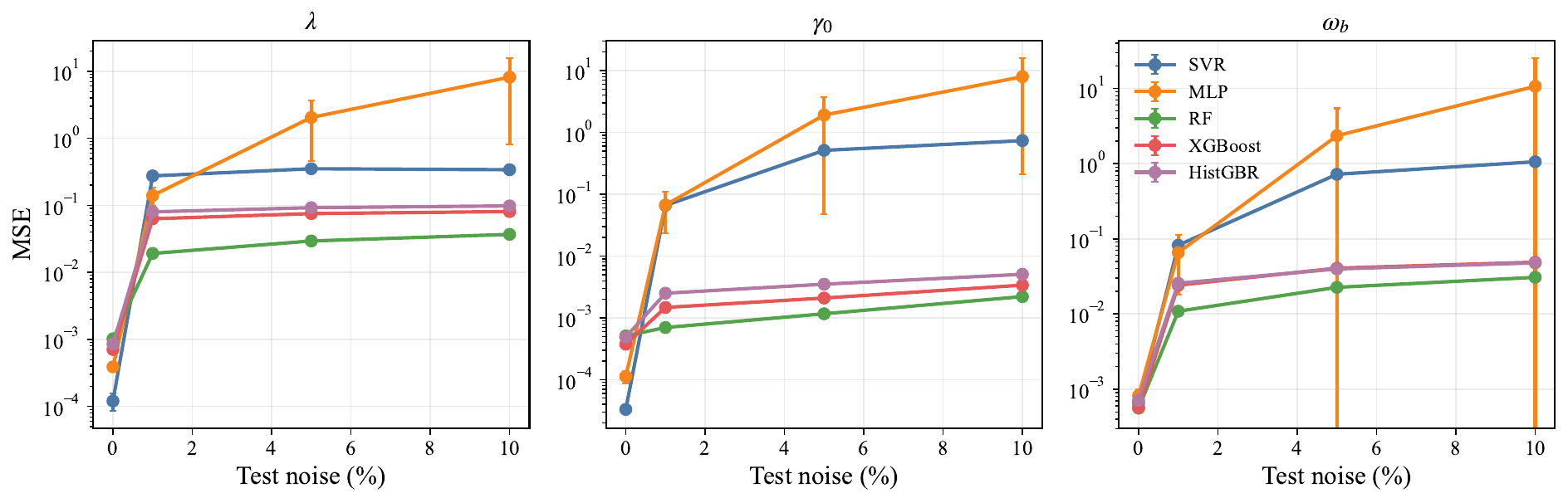}
\caption{Noise robustness of parametric SDF inference in the amplitude-damping (AD) channel. From left to right, the panels show the mean squared error (MSE) for $\lambda$, $\gamma_0$, and $\omega_b$ for models trained on clean signals and tested at $0$, $1$, $5$, and $10\%$ multiplicative noise. Markers and error bars show means and standard deviations over three train--test splits and, at each nonzero noise level, five independent noise realizations. The vertical axes are logarithmic.}

\label{figure1}
\end{figure*}

We study spontaneous emission in the single-excitation manifold, where $\hat{N} =\hat{\sigma}_+\hat{\sigma}_- + \sum_k \hat{a}_k^{\dagger}\hat{a}_k$ is the excitation number operator. Because $[\hat{N},\hat{H}_{\rm AD}]=0$, the state for the initially excited system can be written as $|\Psi(t)\rangle=c_1(t)|1,\mathbf{0}\rangle+\sum_k c_k(t)|0,1_k\rangle$. The initial state is $|\Psi(0)\rangle=|1,\mathbf{0}\rangle$, and the reduced dynamics obeys
\begin{eqnarray}
\dot{\hat{\rho}}_{\rm AD}
&=&-iS(t)[\hat{\sigma}_+\hat{\sigma}_-,\hat{\rho}_{\rm AD}]
\nonumber\\
&&+\gamma(t)\!\left[
\hat{\sigma}_-\hat{\rho}_{\rm AD}\hat{\sigma}_+
-\frac{1}{2}\{\hat{\sigma}_+\hat{\sigma}_-,\hat{\rho}_{\rm AD}\}
\right].
\end{eqnarray}
where $\hat{\rho}_{\rm AD}=\mathrm{Tr}_{\rm b}[|\Psi(t)\rangle\langle\Psi(t)|]$, $S(t)=-\Im[\dot c_1(t)/c_1(t)]$, and $\gamma(t)=-2\Re[\dot c_1(t)/c_1(t)]$~\cite{Breuerbook}. Since populations evolve in this model, a suitable observable is

\begin{eqnarray}
    f_{\rm AD}(t) &=& \langle \hat{\sigma}_z\rangle = 2|c_1(t)|^2-1 \label{AD_signal}, \\
    \dot{c}_1(t) &=& -\int_{0}^{t}K(t-\tau) c_1(\tau) d\tau, \label{c1_dynamics}\\
     K(t) &=&\int_{0}^{\infty} e^{i(\omega_{0}-\omega)t} J(\omega) d\omega.
\end{eqnarray}
Equations~\eqref{PD_signal} and \eqref{AD_signal} define a map
$J(\omega)\mapsto f_m(t)$ with a model label $m \in \{\mbox{PD}, \; \mbox{AD}\}$; throughout, we address the corresponding inverse problem
$f_m(t) \mapsto J(\omega)$. In general, for different open quantum systems the role of the SDF into observables can be derived as is explained in Appendix~\ref{SDF_problem}. Maps in open quantum systems are useful, as we shall illustrate later, for training and guiding optimization in AI-based models, as shown in the schematic overview in Fig.~\ref{Scheme}.

\section{Phenomenological models and parameter estimation}

We introduce parametrized families of SDFs, denoted by $J(\omega,\boldsymbol{\xi})$, with $\boldsymbol{\xi}\in\mathbb{R}^p$ representing model parameters, where $p \geq 1$ is the parameter dimension. In our computational simulations, we use two known SDFs
\begin{eqnarray}
    J(\omega,\boldsymbol{\xi}) &=& 2\alpha \, \omega_c^{\,1-s}\,\omega^{s}\, e^{-\omega/\omega_c} \hspace{0.7 cm} \mbox{PD model},\label{J_Ohmic_param}\\
    J(\omega,\boldsymbol{\xi}) &=& \frac{1}{\pi}\frac{\gamma_0 \lambda^2}{(\omega-\omega_b)^2+\lambda^2} \hspace{0.8 cm} \mbox{AD model}.\label{J_Lorentzian_param}
\end{eqnarray}
where $\boldsymbol{\xi} = [\alpha,s,\omega_c]$ corresponds to the Ohmic class and $\boldsymbol{\xi}=[\gamma_0, \lambda,\omega_b]$ describes Lorentzian functions. Here, $\gamma_0$ is the amplitude, $\lambda$ the width, and $\omega_b$ the bosonic frequency of the peak associated with the Lorentzian SDF. At zero temperature ($T=0$), we obtain $f_{\rm PD}(t) = 2|\rho_{10}(0)|\mathcal{F}_{\rm PD}(t)$ and $f_{\rm AD}(t)= 2\rho_{11}(0)\mathcal{F}_{\rm AD}(t)-1$, where analytical expressions for the signals are given by
\begin{eqnarray}
    \mathcal{F}_{\rm PD}(t) &&= \mbox{exp}\left[-2\alpha \Gamma(s)\int_{0}^{\omega_c t}\frac{\sin\left(s\tan^{-1}x\right)}{(1+x^2)^{s/2}}\,dx\right], \label{Signal_PD}\\
    \mathcal{F}_{\rm AD}(t) &&=  e^{-\lambda t}\left| \cosh\left(\frac{dt}{2} \right)+\frac{\Lambda}{d}\sinh\left(\frac{dt}{2}\right)\right|^2, \label{Signal_AD}
\end{eqnarray}
where $d = [\Lambda^2-4\gamma_0 \lambda]^{1/2}$, $\Lambda = \lambda - i\delta$, $\delta = \omega_0-\omega_b$ is the detuning, and $\Gamma(s)$ is the Gamma function. As in the standard analytically solvable Lorentzian-reservoir model, the pole evaluation of the AD correlation function uses the continuum extension around the resonance. The analytical expression of $\mathcal{F}_{\rm PD}(t)$ is derived in Ref.~\cite{Maniscalco2013}, while the expression for $\mathcal{F}_{\rm AD}(t)$ includes the detuning $\delta$. From Eqs.~\eqref{Signal_PD} and \eqref{Signal_AD}, variations in the model parameter vector $\boldsymbol{\xi}$ leave different, generally correlated signatures in the signal $f_m(t)$. Thus, estimating the SDF from data depends on how sensitively the measured trace responds to $\boldsymbol{\xi}$.

Previous work has quantified the effect of SDF variations on open-system observables~\cite{Plenio2017}. Here, we specialize the discussion to the AD model. Let $J\mapsto J+\delta J$, let $c_1$ and $\widetilde c_1$ denote the corresponding amplitudes, and set $a=\|J\|_1$ and $b=\|\delta J\|_1$, where $\|x\|_1=\int_0^\infty|x(\omega)|d\omega$. Since $|K(t)|\le a$, $|\delta K(t)|\le b$, and physical survival amplitudes satisfy $|c_1(t)|,|\widetilde c_1(t)|\le1$, the difference $u(t)=|\widetilde c_1(t)-c_1(t)|$ obeys the comparison inequality

\begin{equation}
u(t)\le \frac{b t^2}{2}+a\int_0^t(t-\tau)u(\tau)d\tau .
\end{equation}
Solving the corresponding equality gives the rigorous upper bound
\begin{equation}
\big|\Delta f_{\rm AD}(t)\big|
\le \frac{4b}{a}\left[\cosh(\sqrt{a}\,t)-1\right],
\label{eq:AD_sensitivity_bound}
\end{equation}
with the continuous limit $|\Delta f_{\rm AD}(t)|\le2bt^2$ when $a\to0$. 

\begin{figure*}[ht!]
\centering
\includegraphics[width=0.82\linewidth]{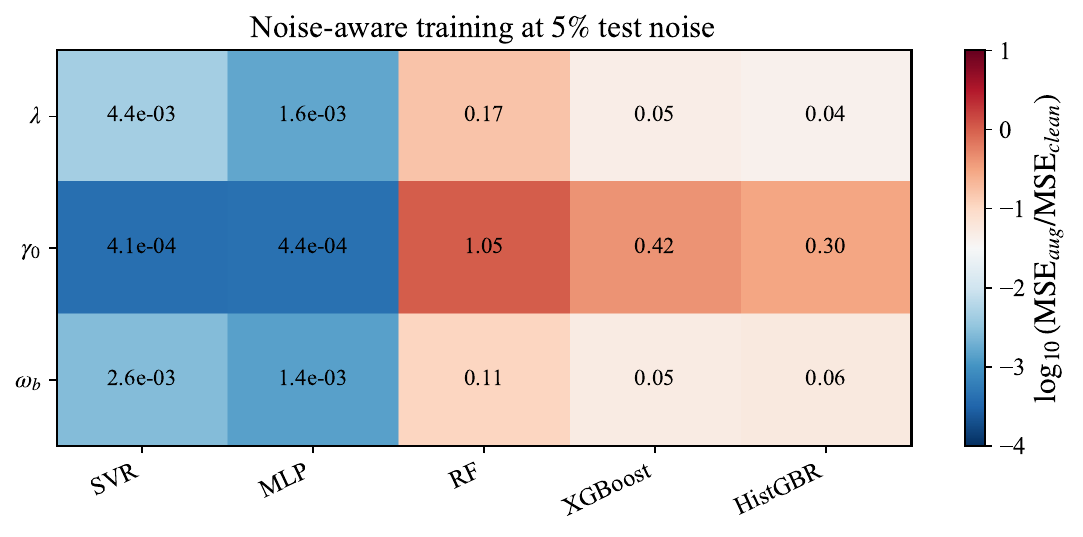}
\caption{Effect of noise-aware training at $5\%$ test noise. Rows correspond to $\lambda$, $\gamma_0$, and $\omega_b$, and columns correspond to the five regression models. Cell annotations give $\mathrm{MSE}_{\rm augmented}/\mathrm{MSE}_{\rm clean}$, while the color scale represents the base-10 logarithm of this ratio. Values below one indicate improvement from noise-aware training. Results are averaged over three train--test splits and five independent test-noise realizations.}
\label{figure4}
\end{figure*}

For the positive-frequency Lorentzian used here,
$a=(\gamma_0\lambda/\pi)[\pi/2+\arctan(\omega_b/\lambda)]$. Integrating Eq.~\eqref{eq:AD_sensitivity_bound} over $t\in[0,t_f]$ gives
\begin{equation}
\int_0^{t_f} |\Delta f_{\rm AD}(t)|dt
\le \frac{4b}{a}\left[
\frac{\sinh(\sqrt{a}\,t_f)}{\sqrt{a}}-t_f\right].
\label{eq:AD_sensitivity_L1time}
\end{equation}
Equation~\eqref{eq:AD_sensitivity_L1time} provides a finite-time worst-case stability bound. It guarantees continuity of the forward map in the $L^1$ norm, while its growth with the observation window also illustrates why inversion becomes progressively sensitive to noise and parameter degeneracy.

\subsection{Simulated data and machine learning regressors}

Simulated data for both PD and AD models are generated by sampling physical regimes in which the signals~\eqref{Signal_PD} and \eqref{Signal_AD} are well defined. To mimic relative measurement fluctuations, we use bounded multiplicative Gaussian noise,
$f_m(t_i)\rightarrow f_m(t_i)[1+\delta_i]$, with
$\delta_i\sim\mathcal{N}(0,\sigma^2)$ independently at each time and clipped to
$[-2\sigma,2\sigma]$. The observable is subsequently clipped to its physical range. Each simulated trace is represented by the vector $\mathbf{x}^{(r)}=[f_m(t_1),\ldots,f_m(t_{N_t})]^{\mathsf T}$ and paired with its generating parameters $\boldsymbol{\xi}^{(r)}$. The supervised dataset is therefore
$\mathcal{D}_m=\{(\mathbf{x}^{(r)},\boldsymbol{\xi}^{(r)})\}_{r=1}^{N_s}$.
For a regressor $R_{\boldsymbol{\theta}}$ that maps a complete time trace to the SDF parameters, training minimizes

\begin{equation}
    \mathcal{L}_{\rm par}(\boldsymbol{\theta})
    =\frac{1}{N_s}\sum_{r=1}^{N_s}
    \left\|R_{\boldsymbol{\theta}}\left(\mathbf{x}^{(r)}\right)
    -\boldsymbol{\xi}^{(r)}\right\|_2^2 .
\end{equation}
In the implementation, one regressor is fitted for each component of $\boldsymbol{\xi}$. The analytical maps~\eqref{Signal_PD} and \eqref{Signal_AD} are used to generate labeled traces, while the trained regressors solve the inverse map directly. A pedagogical description and model hyperparameters are given in Appendix~\ref{ML_appendix}.

For the AD channel, we benchmark support vector regression (SVR), a multilayer perceptron (MLP), random forest (RF), extreme gradient boosting (XGBoost), and histogram-based gradient boosting regression (HistGBR). We generated $6000$ signals on a common grid of $200$ times in $t\in[0,10]$, using $\omega_0=1$ and independently sampling $(\lambda,\gamma_0,\omega_b)$ uniformly from $[0.1,1]$. For each of three random train--test splits, $75\%$ of the traces were used for training and $25\%$ for testing. Models trained on clean signals were evaluated at $0$, $1$, $5$, and $10\%$ test noise, using five independent noise realizations at each nonzero level. In a second protocol, noise-aware training was implemented by assigning each training trace an independent noise standard deviation sampled uniformly from $[0,0.1]$.

Figure~\ref{figure1} shows the mean-square error (MSE) of the different clean-trained models to estimate SDF parameters as a function of the noise amplitude. We observe that RF yields the smallest errors in all cases, and between parameters, $(\lambda,\gamma_0)$ are easier to detect than $\omega_b$. This behavior is consistent with the structure of the forward map: $\lambda$ and $\gamma_0$ jointly control the decay envelope and memory time, whereas $\omega_b$ is encoded through detuning-dependent oscillations that are easily obscured by noise. A finite-difference Jacobian analysis over $1000$ random parameter sets gives median normalized sensitivities $0.379$, $0.415$, and $0.184$ for $\lambda$, $\gamma_0$, and $\omega_b$, respectively. The median Jacobian condition number is $9.17$, with a maximum $1.64\times10^3$, confirming substantial local degeneracies.

Noise-aware training effectively mitigates this failure mode, as demonstrated in Fig.~\ref{figure4}. At $5\%$ test noise, the support vector regression (SVR) mean squared error (MSE) decreases from $0.351$ to $1.56\times10^{-3}$ for $\lambda$, from $0.514$ to $2.11\times10^{-4}$ for $\gamma_0$, and from $0.723$ to $1.88\times10^{-3}$ for $\omega_b$. The corresponding multilayer perceptron (MLP) errors decrease from $2.05$, $1.91$, and $2.35$ to $3.25\times10^{-3}$, $8.39\times10^{-4}$, and $3.22\times10^{-3}$, respectively. Tree ensemble methods also show improvement for the more weakly identifiable $\lambda$ and $\omega_b$ parameters, while random forest (RF) performance changes minimally for $\gamma_0$ due to its inherent robustness. Therefore, the observed degradation is not an intrinsic property of the regressors but rather results from a mismatch between clean training data and noisy observations.

\section{Neural network learning with prior cosine transform}
When the SDF parametrization is unknown, neural networks (NNs) can be used as universal function approximators~\cite{Cybenko1989,Hornik1991}. Using a simulated dataset $\mathcal{D}_m$ and the known quantum map, we can train a deep learning model to infer SDFs from dynamical signals, even in the presence of noise. We now focus on the PD model and the structured SDF

$J(\omega)=J_{\mathrm{bulk}}(\omega)+J_{\mathrm{loc1}}(\omega)+J_{\mathrm{loc2}}(\omega)$~\cite{Ariel2020}, with
\begin{eqnarray}
J_{\mathrm{bulk}}(\omega) &=& 2\alpha\,\omega_c^{1-s}\omega^s e^{-\omega/\omega_c}, \label{Jbulk}\\
J_{\mathrm{loc1}}(\omega) &=& \frac{J_0\omega^s}{\left(\omega/\omega_{\mathrm{loc}}+1\right)^2}
\frac{\Gamma/2}{(\omega-\omega_{\mathrm{loc}})^2+(\Gamma/2)^2}, \label{Jloc1}\\
J_{\mathrm{loc2}}(\omega) &=& J_1\omega^s
e^{-(\omega-\omega_0)^2/(2\sigma^2)}, \label{Jloc2}
\end{eqnarray}
where $s=3$, $\omega_c=1$~THz, $\alpha=0.0275$, $J_1=0.0025$~THz$^{-2}$,
$\sigma=2.4042$~THz, $\omega_0=9.35$~THz,
$J_0=0.0235$~THz$^{-1}$, $\Gamma=0.8414$~THz, and $\omega_{\mathrm{loc}}=15.19$~THz. The structured SDF accurately captures the electron-phonon coupling of the negatively charged silicon-vacancy center in diamond~\cite{Norambuena2016}. However, because of the complexity of the SDF~\eqref{Jbulk}-\eqref{Jloc2}, no closed analytical expression is available for the coherence signal $f_{\rm PD}(t)$; only an approximate expression can be obtained by exploiting the pronounced peak of $J_{\rm loc1}(\omega)$~\cite{Norambuena2016}. The main idea is therefore to find a linear integral transformation between the function $F(f_{\rm PD}(t))$ and the SDF $J(\omega)$, valid for any structured environment. Here, $F(\cdot)$ is a functional operator acting on the signal $f_{\rm PD}(t)$.

In the PD model, the signal $H(t)=-d^2[\ln f_{\mathrm{PD}}(t)]/dt^2$, with $f_{\rm PD}(t)=C(0)\exp[-G(t)]$, satisfies the linear integral relation $H(t)=\int_{0}^{\infty}J(\omega)\coth(\beta\omega/2)\cos(\omega t)\,d\omega$, which has the structure of a cosine transform. Here, the operator is $F(\cdot)=-d^2[\ln(\cdot)]/dt^2$. This motivates introducing the half-range cosine transform $\mathcal{C}\{f\}(\Omega)=\int_{0}^{\infty}f(t)\cos(\Omega t)\,dt$, which is linear because $\mathcal{C}\{\alpha_1 f_1+\alpha_2 f_2\}(\Omega)=\alpha_1\mathcal{C}\{f_1\}(\Omega)+\alpha_2\mathcal{C}\{f_2\}(\Omega)$ for $\alpha_i\in\mathbb{C}$ and functions $f_i(t)$ whose cosine transforms are well defined. The inverse transform is $\mathcal{C}^{-1}\{F\}(t)=(2/\pi)\int_{0}^{\infty}F(\Omega)\cos(\Omega t)\,d\Omega$, which directly yields
\begin{equation} \label{J_from_cosine_add}
J(\Omega)=\frac{2}{\pi}\tanh \left(\frac{\beta \Omega}{2}\right)\,\mathcal{C}\left\{\frac{d^2G}{dt^2}\right\}(\Omega),
\end{equation}
subject to the physical constraint $J(\Omega) \ge 0$. To derive Eq.~\eqref{J_from_cosine_add}, we have used the orthogonality relation $\int_{0}^{\infty}\cos(\omega t)\cos(\Omega t)\,dt=\pi\delta(\Omega-\omega)/2$, where $\delta(x)$ is the Dirac delta. 

\begin{figure}[ht!]
\centering
\includegraphics[width= 1\linewidth]{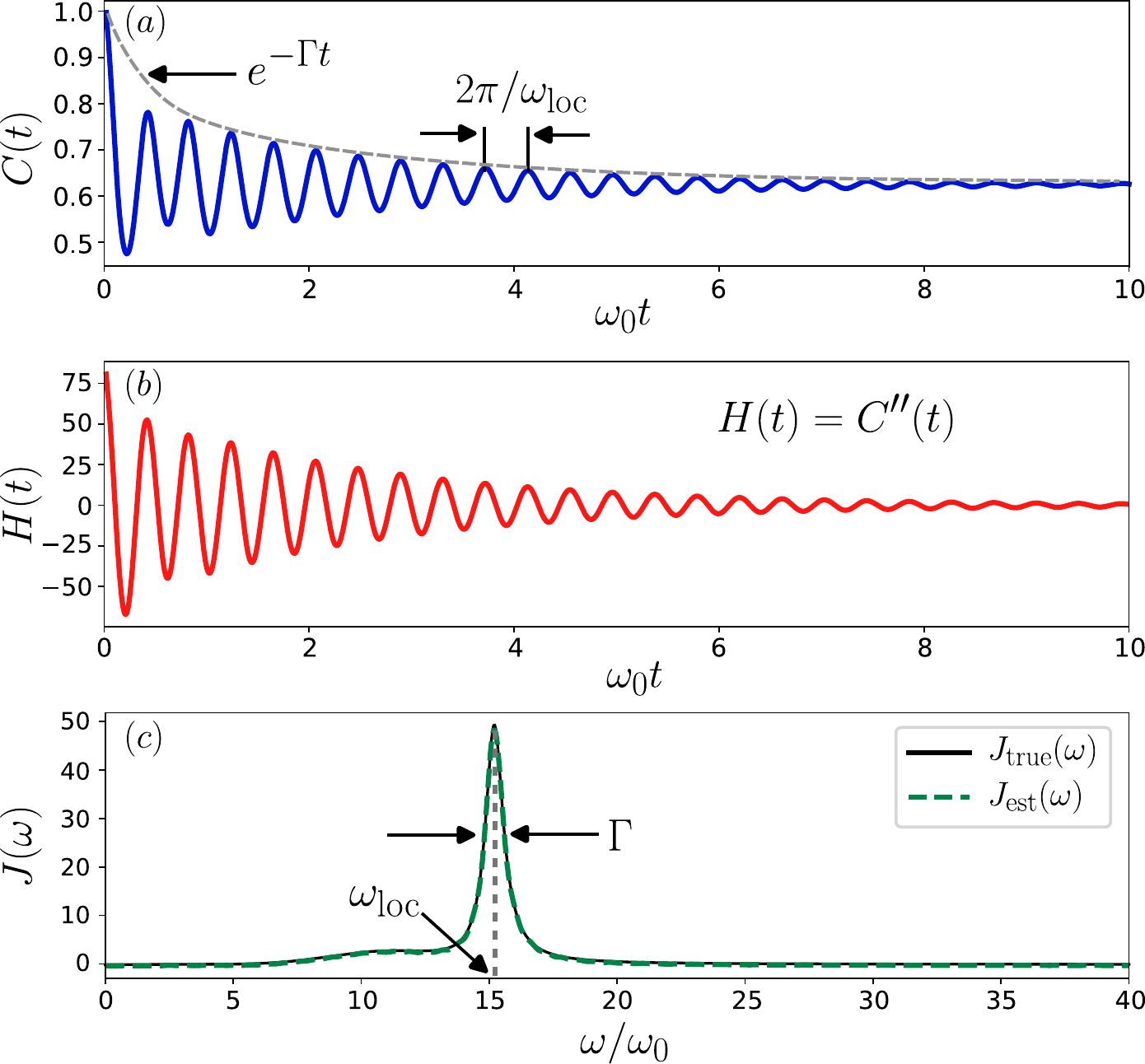}
\caption{Noise-free reconstruction using the discrete cosine transform. (a) Coherence signal generated from the phenomenological SDF in Eqs.~\eqref{Jbulk}--\eqref{Jloc2}. (b) Signal $H(t)=G''(t)$ used in the cosine inversion of Eq.~\eqref{J_from_cosine_add}. (c) True SDF and its discrete-cosine-transform estimate, without neural-network refinement.}
\label{figure2}
\end{figure}

Equation~\eqref{J_from_cosine_add} reveals a theoretical procedure to reconstruct the SDF from $G''(t)=-d^2[\ln f_{\mathrm{PD}}(t)]/dt^2$ and the thermal factor $\tanh(\beta \Omega/2)$. Experimental data, however, consist of discrete noisy time samples. We therefore define an equally spaced time grid $t_n=n\Delta t$ ($n=0,\ldots,N-1$) with $t_f=(N-1)\Delta t$, for which the quadrature $H_{t_f}(\nu_k)=\Delta t\sum_{n=0}^{N-1}H(t_n)\cos(\nu_k t_n)$ with
$\nu_k=\pi k/t_f$ provides a discrete cosine transform estimate. We obtain
\begin{eqnarray} 
H_{t_f}(\nu) &=& \int_{0}^{\infty}J(\omega)\coth\left(\frac{\beta\omega}{2}\right)
K_{t_f}(\nu,\omega)\,d\omega,\\
K_{t_f}(\nu,\omega) &=&
\frac{\sin[(\nu-\omega)t_f]}{2(\nu-\omega)}
+\frac{\sin[(\nu+\omega)t_f]}{2(\nu+\omega)}.
\end{eqnarray}

\begin{figure*}[ht!]
\centering
\includegraphics[width=0.85\linewidth]{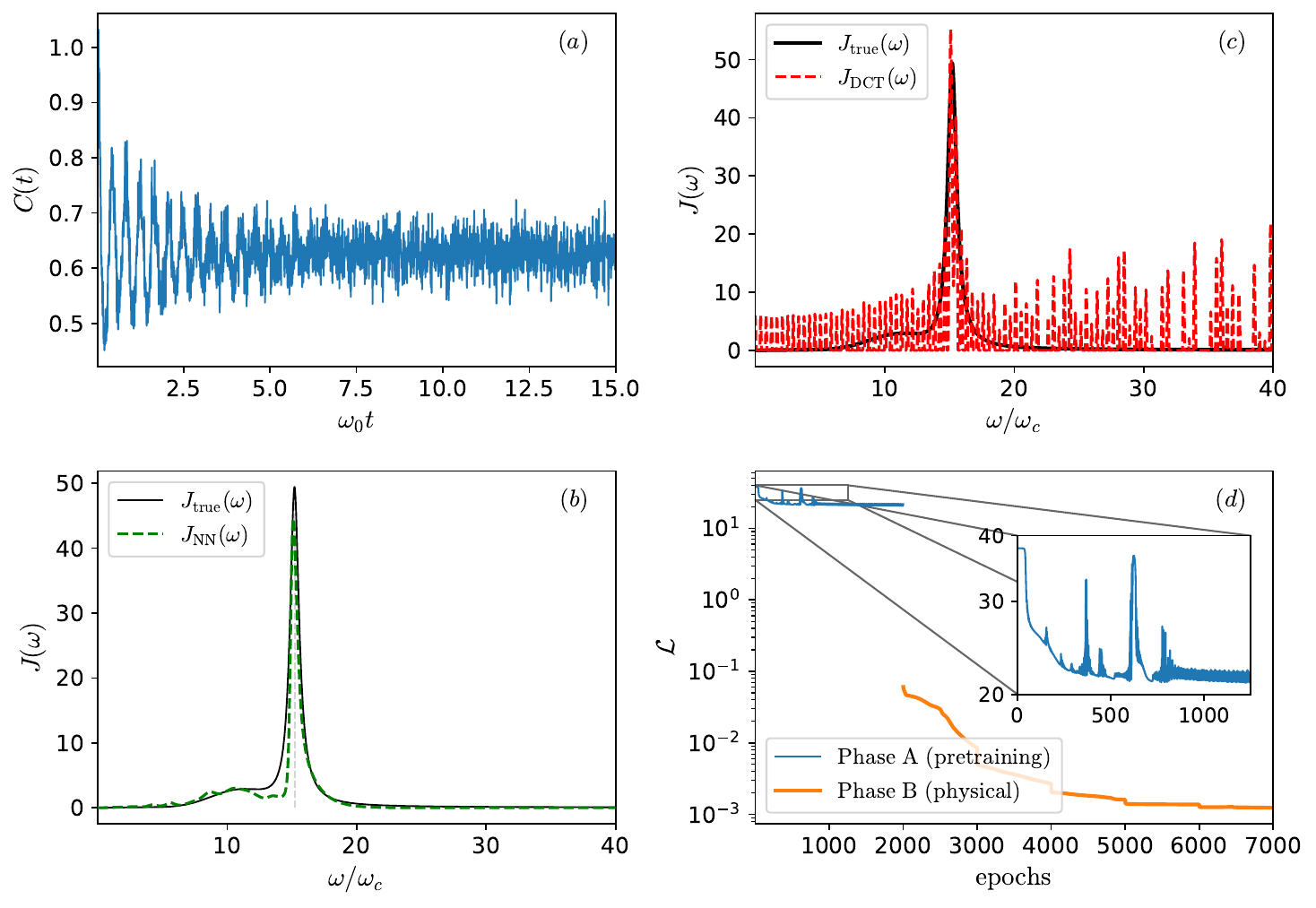}
\caption{Reconstruction of the SDF using a coherence signal with $5\%$ noise. (a) Noisy coherence generated from the phenomenological model. (b) Neural-network reconstruction of the SDF using $J_{\rm DCT}(\omega)$ as the preprocessing prior. (c) True SDF and its discrete-cosine-transform (DCT) estimate. (d) Loss convergence during preprocessing (phase A) and postprocessing (phase B), in which the SDF is refined by the neural network.}

\label{figure3}
\end{figure*}

In the asymptotic limit $t_f \to\infty$ and in the absence of aliasing ($\omega_{\max}<\pi/\Delta t$), this kernel reduces to $H_{t_f}(\nu)\simeq(\pi/2)J(\nu)\coth(\beta\nu/2)$, so the spectral peaks of $H_{t_f}(\nu)$ and $J(\nu)$ coincide up to thermal weighting. These observations motivate a frequency-domain preprocessing stage before applying a deep learning algorithm, but are physically inspired by the model. Naturally, for a different open quantum system, a linear integral relation $H(t) \mapsto J(\omega)$ must be derived, taking into account specific details of the model. We then define the dataset $\mathcal{D}_{\omega}\equiv\{\mathcal{S}(\omega_n),\omega_n\}$ with $\mathcal{S}(\omega_n)=[2/(\pi\coth(\beta\omega_n/2))]H_T(\omega_n)$. This provides a coarse reconstruction of $J(\omega)$ and a starting point for neural network (NN) optimization. The NN is first pre-trained by minimizing the loss function (phase A)
\begin{equation}
\mathcal{L}_{\mathrm{pre}}
=\frac{1}{L}\sum_{n=1}^{L}\!
\big[J_{\mathrm{NN}}(\omega_n;\boldsymbol{\theta})-\mathcal{S}(\omega_n)\big]^2,
\end{equation}
yielding optimal NN parameters $\boldsymbol{\theta}^\ast$ after training and fine tuning. In a second postprocessing stage, the reconstruction of the SDF is refined using time domain information encoded in the signal $f_m(t)$, by minimizing the loss function (phase B)
\begin{equation}
\mathcal{L}_{\mathrm{post}}
=\frac{1}{N}\sum_{i=1}^{N}\!
\big[\hat{f}_m(t_i,J_{\mathrm{NN}}(\omega;\boldsymbol{\varphi}))
-\left[f_m(t_i)\left(1+\delta_i \right)\right]\big]^2,
\end{equation}
initialized at $\boldsymbol{\varphi}=\boldsymbol{\theta}^\ast$ obtained in the pretraining phase. Here, $\hat{f}_m(t_i,J_{\mathrm{NN}}(\omega;\boldsymbol{\varphi}))$ is computed from the known quantum map $J(\omega) \mapsto f_m(t)$. Physical consistency is enforced by parametrizing

\begin{equation}
J_{\mathrm{NN}}(\omega;\boldsymbol{\varphi})
=g\,\mathcal{F}(\omega)\,
\operatorname{softplus}[\mathbf{N}_J(\omega;\boldsymbol{\varphi})],
\end{equation}
where $g>0$ is a learned gain, $\mathcal{F}(\omega)=\omega^d e^{-\omega/\Omega}$ is normalized by its maximum on the numerical grid, and $\operatorname{softplus}(x) = \ln(1+e^{x})$ is the softplus function used in machine learning. This parameterization guarantees non-negativity and imposes the required low- and high-frequency behavior. More details of the NN protocol are given in Appendix~\ref{NN_protocol}.

\section{Results of SDF reconstruction for the PD model using neural networks refinement}
In the ideal noise-free setting (Fig.~\ref{figure2}), the cosine transform in Eq.~\eqref{J_from_cosine_add} already provides an accurate and parameter-free SDF reconstruction: it reproduces both the super-Ohmic envelope and the dominant localized features of $J(\omega)$ at negligible computational cost. This establishes a useful baseline: whenever the model is well known and the signal-to-noise ratio is high, a linear inversion can provide a good approximation to the SDF without imposing a phenomenological SDF. The present formulation therefore broadens previous approaches~\cite{Barr2024,Barr2025,Barr2025_arxiv}, which focused on noise-free scenarios and phenomenological SDF families.

\begin{figure*}[ht!]
\centering
\includegraphics[width=0.88\linewidth]{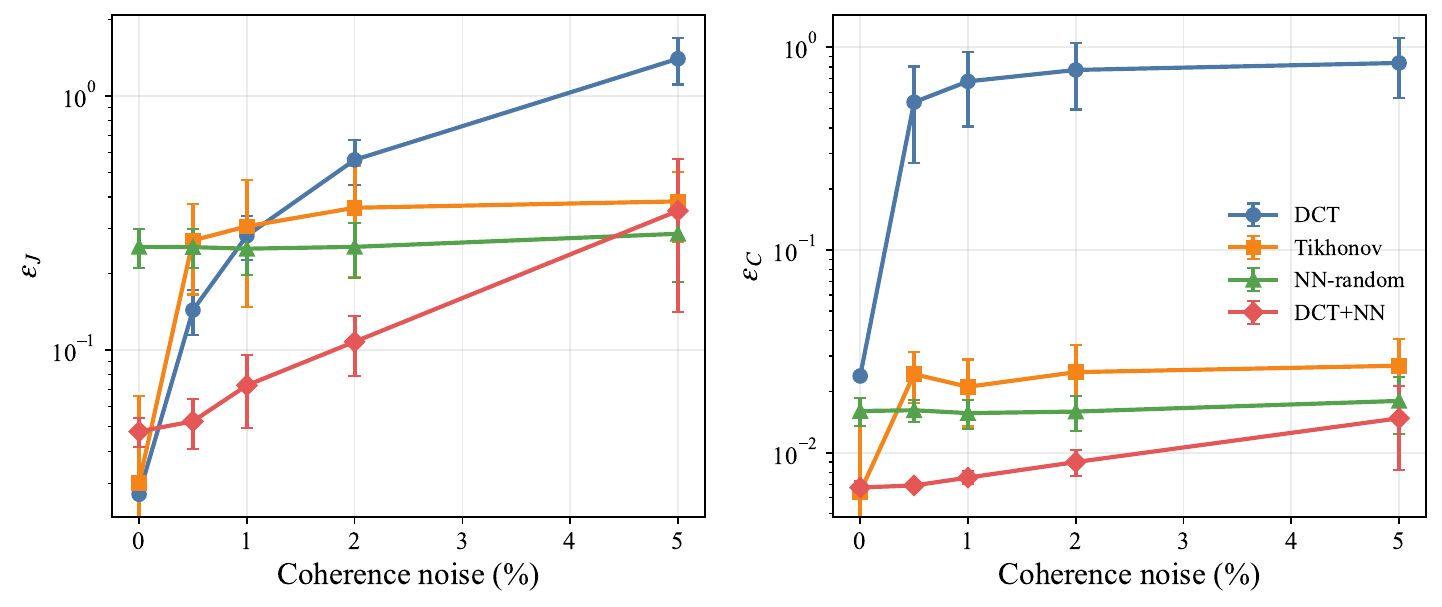}
\caption{Quantitative ablation of the pure-dephasing reconstruction. (a) Relative spectral error $\varepsilon_J$ and (b) coherence error $\varepsilon_C$ versus multiplicative coherence noise for the raw DCT, nonnegative Tikhonov inversion, randomly initialized constrained neural network (NN), and DCT-pretrained constrained NN. Markers and error bars show means and standard deviations over five independent noise and training seeds. Both vertical axes are logarithmic.}
\label{figure5}
\end{figure*}

The situation changes in the presence of noise. We perturb the coherence as $C(t)\rightarrow C(t)[1+\delta(t)]$, producing the signal observed in Fig.~\ref{figure3}(a). The discrete cosine transform then yields a nonphysical SDF with fictitious oscillations and negative regions, as shown in Fig.~\ref{figure3}(c). This is expected because the second derivative in $H(t)=-d^2[\ln f_{\mathrm{PD}}(t)]/dt^2$ amplifies short-time fluctuations and finite-window artifacts. We mitigate this instability by combining the cosine-based spectral prior (phase A) with a physics-constrained NN refinement (phase B). To quantify the SDF reconstruction, we use the error metrics
\begin{equation}
\varepsilon_J=\frac{\|J_{\rm rec}-J_{\rm true}\|_2}{\|J_{\rm true}\|_2},
\qquad
\varepsilon_C=\frac{\|C_{\rm rec}-C_{\rm true}\|_2}{\|C_{\rm true}\|_2},
\end{equation}
together with the dominant-peak position and full-width-at-half-maximum errors and the fraction of negative spectral weight. Figure~\ref{figure5} reports means and standard deviations over five independent noise and training seeds. We compare the raw DCT, a nonnegative second-order Tikhonov inversion with its regularization strength selected on held-out time points~\cite{Hansen1998}, the constrained NN trained from random initialization, and the full DCT+NN protocol presented in this work.

In the noise-free limit, the DCT gives $\varepsilon_J=0.027$, and additional NN processing is unnecessary. At $0.5$, $1$, and $2\%$ noise, the raw-DCT errors increase to $0.144\pm0.029$, $0.282\pm0.056$, and $0.560\pm0.112$, whereas the full DCT+NN errors remain $0.053\pm0.012$, $0.073\pm0.023$, and $0.108\pm0.028$. The NN trained from random initialization remains near $\varepsilon_J\simeq0.25$, demonstrating the value of spectral pretraining in this moderate-noise regime. At $5\%$ noise, the DCT fails ($\varepsilon_J=1.40\pm0.29$ and a negative-weight fraction $0.52\pm0.19$). Both NN variants regularize the inversion; their global spectral errors overlap within the run-to-run variation, but the DCT-pretrained NN has a smaller coherence error ($0.0148$ versus $0.0180$) and a much smaller relative peak-width error ($0.080$ versus $0.648$). The classical Tikhonov baseline is competitive, especially in the noise-free limit, but at $1$-$2\%$ noise, its mean spectral error is approximately three to four times larger than that of DCT+NN. The NN should therefore be viewed as an adaptive physics-constrained regularizer, not as a universal replacement for classical inverse methods.

\begin{figure}[ht!]
\centering
\includegraphics[width=\linewidth]{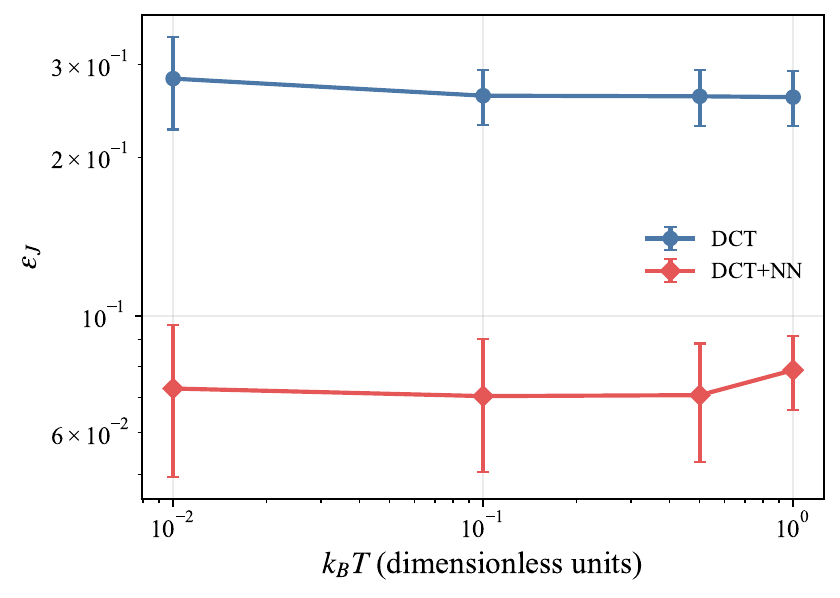}
\caption{Relative spectral error $\varepsilon_J$ versus the known dimensionless temperature $k_B T$ at $1\%$ coherence noise for the raw DCT and DCT-pretrained neural-network reconstructions. Markers and error bars show means and standard deviations over three independent seeds. Both axes are logarithmic.}
\label{figure6}
\end{figure}

Temperature enters through $\coth(\beta\omega/2)$ in the forward map and $\tanh(\beta\omega/2)$ in the inversion. Figure~\ref{figure6} shows that, for known temperature and the structured spectrum considered here, the DCT+NN spectral error remains between $0.070$ and $0.079$ over $k_BT\in[0.01,1]$, while the DCT error remains near $0.26$-$0.28$. If $\beta$ is not independently calibrated, however, it becomes partially degenerate with the low-frequency behavior of $J(\omega)$ and should be inferred jointly using additional observables or informative priors.

These results show that the proposed framework regularizes the ill-conditioned inversion within the tested model class and controlled noise range; it does not remove the intrinsic ill-conditioning. The preprocessing remains channel specific. For PD, the logarithm and second derivative expose a linear cosine relation. For AD, $J(\omega)$ enters a memory kernel and then a nonlinear Volterra equation, so there is no equally direct DCT prior. Parametric regression can nevertheless be used for PD when a reliable SDF family is known, and a nonparametric NN can be used for AD by differentiating through a Volterra solver. Broadband backgrounds may also be combined with parametric components while localized residual peaks are represented by a constrained NN or sparse basis. Experimental applications will additionally require calibrated colored-noise, state-preparation and measurement, and model-mismatch terms in the likelihood.

\section{Conclusions}
We developed and benchmarked two complementary, data-driven routes to reconstruct spectral density functions in exactly solvable open quantum models. Inverting time-domain signals to obtain frequency-dependent SDFs is an ill-conditioned problem. Small perturbations in the data can lead to large deviations in the reconstructed spectra. This behavior is expected for compact integral operators (see Ref.~\cite{Plenio2017} for related bounds). 

In the parametric regime, regression techniques provide fast, interpretable estimates of SDF parameters, but their apparent noise robustness depends strongly on whether the training distribution includes measurement fluctuations. Noise augmentation reduces the largest AD regression errors by orders of magnitude. This route nevertheless assumes prior knowledge of the phenomenological family. In the nonparametric regime, the cosine-transform structure of the quantum map induces a physics-consistent spectral prior when the signal-to-noise ratio is sufficiently high. A constrained neural-network refinement then acts as a structured regularization mechanism, with a clear advantage over raw DCT and random initialization at low-to-moderate noise and a more limited, feature-dependent advantage at $5\%$ noise.

From a mathematical perspective, the quantum map between dynamical observables and the SDF involves integral relations that define compact operators in frequency space. Their inversion amplifies high-frequency components and is intrinsically ill-conditioned under finite-time sampling and noisy data. The cosine prior, Tikhonov baseline, and constrained NN should therefore be understood as alternative regularization mechanisms with different bias--variance tradeoffs. The present results provide a reproducible benchmark for environmental spectroscopy of structured reservoirs; validation with real experimental noise and model mismatch remains an essential next step.

\section*{Data and code availability}
The data and code supporting the findings of this article are publicly
available in Ref.~\cite{CodeRepository2026}. The repository contains the
source code, benchmark data, fixed random seeds, raw per-run metrics,
aggregate tables, figure data, software requirements, and step-by-step
instructions needed to reproduce the results.

\section*{Acknowledgments}
F.P., J.V.S.K.L., P.M.P. and F.F.F. acknowledge support from Funda\c{c}\~{a}o de Amparo \`{a} Pesquisa do Estado de S\~{a}o Paulo (FAPESP), project numbers 2024/20249-8, 2024/15741-0, 2025/08578-9 and 2023/04987-6. A.N.\ acknowledges support from Fondecyt Regular No.~1251131, Fondecyt Exploraci\'on No.~13250014, and Anillo Tem\'atico ATE 250066.

\appendix

\section{GKSL derivation and dependence on the SDF}\label{SDF_problem}

The starting point of the theory of open quantum systems is the Hamiltonian of the full (closed) system
\begin{equation}
    \hat{H} = \hat{H}_s \otimes \mathds{1}_b + \mathds{1}_s \otimes \hat{H}_b + \hat{V},
\end{equation}
with $\hat{H}_s = \sum_a \hbar \omega_a \ket{a}\!\bra{a}$, $\hat{H}_b = \sum_k \hbar \omega_k \hat{a}_{k}^{\dagger}\hat{a}_k$, and $\hat{V} = \hat{S}\hat{X}^{\dagger} + \hat{S}^{\dagger}\hat{X}$. Here, we are only considering a collective bath operator of the form $\hat{X} = \sum_k g_k \hat{a}_k$, where $g_k$ encodes the system-bath couplings. For initially uncorrelated states $\hat{\rho}(0) = \hat{\rho}_s(0) \otimes \hat{\sigma}_b$, we assume a bath density matrix in thermal equilibrium
\begin{equation}
    \hat{\sigma}_b = \frac{e^{-\beta \hat{H}_b}}{\mathrm{Tr}[e^{-\beta \hat{H}_b}]}, \qquad \beta = (k_B T)^{-1},
\end{equation}
which maximizes the von Neumann entropy for a fixed average energy. Under the weak-coupling and secular approximations, the reduced state $\hat{\rho}_s(t) = \mathrm{Tr}_b[\hat{\rho}(t)]$ follows the Gorini-Kossakowski-Sudarshan-Lindblad (GKSL) equation ($\hbar=1$),

\begin{eqnarray}
    \dot{\hat{\rho}}_s &=&
    -i[\hat{H}_s+\hat{H}_{\rm LS}(t),\hat{\rho}_s(t)]
    \nonumber\\
    &&+\sum_{\Omega,n}\gamma_{n}(t)
    \Big[\hat{S}_{n}(\Omega)\hat{\rho}_s(t)
    \hat{S}^{\dagger}_{n}(\Omega) \nonumber\\
    &&\hspace{1.3cm}
    -\tfrac{1}{2}
    \{\hat{S}^{\dagger}_{n}(\Omega)\hat{S}_{n}(\Omega),
    \hat{\rho}_s(t)\}\Big].
\end{eqnarray}
Here, $\hat{H}_{\rm LS}(t)$ denotes the Lamb-shift correction, and the system transition operators are

\begin{equation}
    \hat{S}_n(\Omega)
    =\sum_{\omega_b-\omega_a=\Omega}
    \ket{a}\bra{a}\hat{S}_n\ket{b}\bra{b}.
\end{equation}
For a positive Bohr frequency $\Omega$, the second-order time-local absorption and emission rates have the transparent form
\begin{eqnarray}
\gamma_{\uparrow}(\Omega,t)&=&2\int_0^\infty J(\omega)n(\omega)
\frac{\sin[(\omega-\Omega)t]}{\omega-\Omega}\,d\omega,\\
\gamma_{\downarrow}(\Omega,t)&=&2\int_0^\infty J(\omega)[n(\omega)+1]
\frac{\sin[(\omega-\Omega)t]}{\omega-\Omega}\,d\omega,
\end{eqnarray}
where the ratio is understood by continuity at $\omega=\Omega$. Thus, the reduced density matrix naturally depends on $J(\omega)$, motivating the inverse problem of reconstructing the SDF from observables $\langle\hat{\mathcal O}\rangle=\mathrm{Tr}_s[\hat{\rho}_s(t)\hat{\mathcal O}]$.

\section{Machine-learning models for parameter estimation}\label{ML_appendix}

\subsection{Learning task in plain language}

For the parametric method, a single input example is an entire population trace sampled at the same $N_t=200$ times,

\begin{equation}
\mathbf{x}=[f_{\rm AD}(t_1),\ldots,f_{\rm AD}(t_{N_t})]^{\mathsf T}.
\end{equation}
The desired output is one physical parameter of the Lorentzian reservoir, namely $\lambda$, $\gamma_0$, or $\omega_b$. We train three separate regressors for each model family, one per target. This is analogous to learning three calibrated measuring instruments: each instrument reads the same time trace but is optimized to report a different physical quantity. Input standardization is used for SVR and the MLP, whereas tree models operate directly on the sampled signal values.

\subsection{Why several regressors are compared}

\paragraph{Support vector regression.}
SVR predicts a parameter from a weighted combination of representative training traces~\cite{Cortes1995}. Its radial-basis kernel can describe smooth nonlinear relations while the $\epsilon$-insensitive loss ignores sufficiently small residuals. This often yields excellent accuracy for clean, low-dimensional inverse maps, but a perturbation distributed across all time samples can change many kernel distances simultaneously.

\paragraph{Multilayer perceptron.}
The MLP is a feed-forward neural network that repeatedly applies affine maps and nonlinear activation functions. It learns a global smooth approximation to the inverse map and can combine information from the decay envelope and oscillatory features. Its flexibility is useful, but without representative noise during training it may interpret measurement fluctuations as physical structure.

\paragraph{Random forest.}
RF averages many decision trees trained on bootstrap samples and randomly selected subsets of input features~\cite{Breiman2001}. Each tree partitions the space of time traces into regions with similar target values. Averaging reduces variance and makes the model comparatively insensitive to isolated distortions of individual time points.

\paragraph{Gradient-boosted trees.}
XGBoost and HistGBR build trees sequentially, with each new tree correcting residual errors left by the previous ensemble~\cite{Friedman2001,Chen2016}. XGBoost uses explicit row and feature subsampling together with regularization. HistGBR first bins continuous inputs into histograms, which makes training efficient and introduces an additional coarse-graining that can be helpful in noisy settings.

\subsection{Exact implementation and hyperparameters}

The benchmark uses $N_s=6000$ traces with $(\lambda,\gamma_0,\omega_b)$ independently sampled from $[0.1,1]$, a $75/25$ train--test split, three independent split seeds, and five independent noise realizations for every nonzero test-noise level. Noise-aware training assigns each training trace a standard deviation drawn uniformly from $[0,0.1]$. The exact estimators are:
\begin{itemize}
\item SVR: radial basis kernel, $C=30$, $\epsilon=0.01$, and \texttt{gamma=scale}, preceded by feature standardization.
\item MLP: hidden widths $(192,96)$, ReLU activations, Adam optimizer, $L^2$ penalty $10^{-4}$, initial learning rate $8\times10^{-4}$, at most $450$ iterations, a $15\%$ internal validation subset, and early stopping after $30$ non-improving iterations; inputs are standardized.
\item RF: $350$ trees, $70\%$ of features considered at each split, one minimum sample per leaf, bootstrap aggregation, and all CPU cores.
\item XGBoost: $550$ trees of maximum depth $6$, learning rate $0.045$, row subsampling $0.85$, column subsampling $0.8$, $L^2$ regularization $1$, squared-error objective, and histogram tree construction.
\item HistGBR: at most $350$ boosting iterations, learning rate $0.07$, at most $31$ leaves per tree, $L^2$ regularization $10^{-3}$, and automatic early stopping.
\end{itemize}
Performance is reported using
\begin{eqnarray}
\mathrm{MSE}_q&=&\frac{1}{N_{\rm test}}\sum_r
(\hat{\xi}_{q}^{(r)}-\xi_q^{(r)})^2,\\
\mathrm{MAE}_q&=&\frac{1}{N_{\rm test}}\sum_r
|\hat{\xi}_{q}^{(r)}-\xi_q^{(r)}|,
\end{eqnarray}
for $q\in\{\lambda,\gamma_0,\omega_b\}$. The random seeds, fitted configurations, raw predictions summarized by run, and plotting routines are provided in Ref.~\cite{CodeRepository2026}.

\section{Algorithmic protocol for reconstructing spectral density functions using cosine transform} \label{NN_protocol}

The reported PD benchmark uses $N=800$ equally spaced times on $t\in[0,15]$ and $M=400$ frequencies on $\omega\in[10^{-3},40]$. More precisely, the full coherence signal is generated on the $N=800$-point time grid, while a uniformly subsampled set of $500$ time points is used in the optimization losses to reduce computational cost. We set $\rho_{eg}(0)=0.5$, use single-precision tensors, and evaluate the frequency integrals with trapezoidal weights. Independent random seeds $0,\ldots,4$ generate the noise and initialize the networks. The temperature sweep uses seeds $0,1,2$. The neural-network stage should be understood as a physics-constrained reconstruction scheme: physical information is imposed through the positivity-preserving output, the asymptotic filter, and the forward open-system map, rather than through a differential-equation residual as in conventional physics-informed neural networks.

\paragraph{Signal and cosine prior.}
The noisy coherence is clipped to $[10^{-8},1]$ before computing
$G(t)=-\ln[C(t)/(2\rho_{eg}(0))]$. First and second derivatives are evaluated with second-order finite differences, and Eq.~\eqref{J_from_cosine_add} is integrated by the trapezoidal rule. The raw DCT is retained for the reported baseline and negativity metric. For NN pretraining only, negative values are set to zero, positive outliers are capped at the $99.5$th percentile, and an 11-point third-order Savitzky--Golay filter is applied. Thus, the DCT output plays two distinct roles: the unmodified DCT is used as a diagnostic baseline, whereas a clipped and smoothed version is used only as a stable spectral prior for Phase A.

\paragraph{Network architecture.}
$\mathbf{N}_J(\omega)$ is a fully connected feed-forward network with one scalar input, four hidden layers of $128$ neurons, hyperbolic-tangent activations, and one scalar output. Including the learned positive gain, the model has $49\,922$ trainable parameters. We use
$\mathcal{F}(\omega)=\omega^3\exp(-\omega/\Omega)$ with $\Omega=4$, normalized by its maximum on the frequency grid, and a softplus output to enforce positivity. The factor $\mathcal{F}(\omega)$ fixes the expected low-frequency scaling and suppresses unphysical high-frequency spectral weight, while the neural network learns the remaining structured frequency dependence.

More explicitly, with normalized input $z_0=2\omega/\omega_{\max}-1$, the network evaluates
\begin{eqnarray}
\mathbf{z}_{\ell}&=&\tanh(\mathbf{W}_{\ell}\mathbf{z}_{\ell-1}+\mathbf{b}_{\ell}),
\qquad \ell=1,\ldots,4,\\
r(\omega)&=&\mathbf{W}_5\mathbf{z}_4+\mathbf{b}_5,\\
\mathcal{F}(\omega)&=&
\frac{\omega^3e^{-\omega/4}}
{\max_{\omega_n}(\omega_n^3e^{-\omega_n/4})},\\
J_{\rm NN}(\omega)&=&
\operatorname{softplus}(g_{\rm raw})\,
\mathcal{F}(\omega)\operatorname{softplus}[r(\omega)].
\end{eqnarray}
The layer dimensions are $1\!\to\!128\!\to\!128\!\to\!128\!\to\!128\!\to\!1$. All weights and biases, together with the scalar $g_{\rm raw}$, are trainable. PyTorch default linear-layer initialization is used after fixing the stated random seed.

\paragraph{Optimization.}
Let $\Delta^2J_n=J_{n+1}-2J_n+J_{n-1}$ and $S_J=\max(10^{-6},\max_nJ_n)$. The scale-invariant curvature penalty is
\begin{equation}
\mathcal{R}_{\rm curv}
=\frac{1}{M-2}\sum_{n=2}^{M-1}
\left(\frac{\Delta^2J_n}{S_J}\right)^2.
\end{equation}
Phase A uses AdamW with learning rate $2\times10^{-3}$, weight decay $10^{-6}$, and at most $350$ epochs to minimize
\begin{equation}
\mathcal{L}_{A}
=\operatorname{SmoothL1}\!\left(
\frac{J_{\rm NN}}{S_{\rm prior}},
\frac{J_{\rm prior}}{S_{\rm prior}}\right)
+2\times10^{-5}\mathcal{R}_{\rm curv},
\end{equation}
where $S_{\rm prior}=\max(1,\max_nJ_{{\rm prior},n})$. 
Here, $\operatorname{SmoothL1}$ denotes the mean smooth-$L_1$ loss over the frequency grid, using the default PyTorch transition parameter.
Phase B uses AdamW with learning rate $7\times10^{-4}$ and at most $700$ epochs to minimize
\begin{equation}
\begin{aligned}
\mathcal{L}_{B}={}&
\frac{1}{N_{\rm tr}}\sum_i
\left[\frac{C_{\rm NN}(t_i)-C_{\rm obs}(t_i)}{S_C}\right]^2 \\
&+2\times10^{-5}\mathcal{R}_{\rm curv} \\
&+\frac{10^{-2}d_e}{M S_A^2}
\sum_n\left(J_{{\rm NN},n}-J_{A,n}\right)^2 .
\end{aligned}
\end{equation}

where $S_C=\max[10^{-4},\sqrt{\langle C_{\rm obs}^2\rangle}]$, $S_A=\max(1,\max_nJ_{A,n})$, and $d_e=\max[0.1,1-e/(699)]$ decays with epoch $e$. The quantity $J_A$ denotes the spectral reconstruction obtained at the end of Phase A; the last term in $\mathcal{L}_B$ therefore acts as a weak, epoch-decaying prior that stabilizes the early stage of the time-domain refinement without forcing the final solution to remain fixed at the DCT-pretrained spectrum.
The learning rate is halved after $80$ epochs without improvement, with floor $2\times10^{-5}$. Early stopping is triggered after $100$ and $160$ non-improving epochs in phases A and B, respectively. Gradients are clipped at Euclidean norm $5$.

\paragraph{Baselines and diagnostics.}
The random-initialization ablation uses the same architecture, physical filter, phase-B loss, and stopping criteria but omits phase A and the prior term. The Tikhonov baseline minimizes the time-domain residual with non-negativity and a second-difference penalty; its strength is selected from $\{10^{-5},10^{-4},10^{-3},10^{-2},10^{-1},1\}$ using a held-out $20\%$ subset of the time samples. All methods are evaluated using $\varepsilon_J$, $\varepsilon_C$, spectral $R^2$, dominant-peak position and width, negative spectral weight, and runtime. Complete scripts, seeds, raw per-run metrics, aggregate tables, figure data, and environment specifications are available in Ref.~\cite{CodeRepository2026}. These diagnostics are intended to quantify both spectral accuracy and dynamical consistency: $\varepsilon_J$ and the peak metrics assess the reconstructed SDF itself, whereas $\varepsilon_C$ measures whether the learned spectrum reproduces the observed open-system coherence.


\end{document}